# Femtosecond pulse with THz repetition frequency based on strong coupling between quantum emitters and a plasmonic resonator

Shi-Lei Li, Rongzhen Jiao, Li Yu

*State Key Laboratory of Information Photonics and Optical Communications, School of Science, Beijing University of Posts and Telecommunications, Beijing 100876, China*

Nanoscale pulsed light is highly desirable in nano-integrated optics. In this letter, we obtained femtosecond pulses with THz repetition frequency via the strong coupling between quantum emitters (QEs) and plasmonic resonators. Our structure consists of a V-groove (VG) plasmonic resonator and a nanowire embedded with two-level QEs. The influences of the incident light intensity and QE number density on the transmission response for this hybrid system are investigated through semi-classical theory and simulation. It is found that the transmission response can be modulated to the pulse form. The reason is the strong coupling causes the output power of nanowire to behave as an oscillating form, the oscillating output power in turn causes the field amplitude in the resonator to oscillate over time. A feedback system is formed between the plasmonic resonator and the QEs in nanowire. This provides a new method for generating narrow pulsed lasers with ultrahigh repetition frequency in plasmonic systems using a continous wave input, which has potential applications in generating optical clock signals at the nanoscale.

## I. INTRODUCTION

In recent years, the coupling (weak and strong coupling) between QEs and plasmonic resonators has attracted much attention because of its various applications in quantum information [1,2], thresholdless lasing [3] and photon detectors [4]. We find that the strong coupling between QEs and plasmonic resonators can also be used to generate femtosecond pulses with ultrahigh repetition frequency in the case of a continuous wave input. Small on-chip pulsed lasers exhibiting mode sizes smaller than the emission wavelength represent an ideal solution for light source integration [5,6]. Plasmon pulsed lasers are especially promising for confining the emission below the diffraction limit while also exhibiting improved performance provided by enhanced emission dynamics [7-10]. Plasmonic waveguide-resonator (PWR) systems which guide radiation in the form of strongly confined surface plasmon-polariton modes represent a promising solution to manipulate light field at the nanoscale [11-17]. Fundamentally, strong localization of field can enhance the energy exchange of QEs with light [18,19]. An attractive type of PWR is VG waveguide-resonator system, which can strongly confine electromagnetic field at the nanoscale and is promising for developing a planar plasmonic circuitry platform [20-24]. To the best of our knowledge, in the case of a continuous wave input, femtosecond pulses with THz repetition frequency are rarely reported in plasmonic resonator systems. Therefore, there is an increasing demand for femtosecond pulsed lasers with ultrahigh repetition frequency at the nanoscale.

In this letter, we investigate the influences of the incident light intensity and the QE number density on the transmission response for the hybrid system composed of a plasmonic VG resonator coupled to a nanowire embedded with QEs. In the weak excitation limit, Rabi splitting occurs in the transmission spectrum of the system, demonstrating that the coupling between VG resonator and the QEs in nanowire is strong coupling. When the intensity of the excited light is strong, the hybrid system is no longer in the weak excitation limit, the QEs will oscillate between the upper and lower energy levels, thereby causing the output power of the nanowire to oscillate in pulse form, and then the transmittance of this system also oscillates with time in pulse form. Moreover, it is found that the repetition frequency and extinction ratio of the pulses can be freely controlled by the incident light intensity and the QE number density. This provides a new method for generating narrow pulsed lasers with ultrahigh repetition frequency in plasmonic resonator systems.

The structure shown in Fig. 1 consists of two plasmonic waveguides ($S_1$ and $S_2$), a VG cavity and a nanowire embedded with QEs. The dielectric around the VG cavity is silver, the dielectric constant of which is obtained from the experimental data of Johnson and Christ [25]. The transmittance of the system is numerically calculated using finite difference time domain (FDTD) method. In order to excite the SPPs, the input light is set to be continuous wave (CW) of transverse magnetic (TM) mode. The complex dielectric function of the nanowire embedded with QEs is described by a Lorentz model [26]:
$\varepsilon(\omega) = \varepsilon_\infty - f\omega_L^2 / (\omega^2 - \omega_L^2 + i\gamma\omega)$.

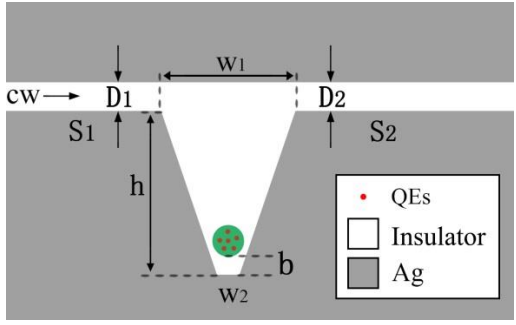

FIG. 1. Schematic diagram of the hybrid system of a plasmonic VG resonator coupled to a nanowire embedded with QEs. The refractive index of the insulator is 1.41. The widths of waveguides $S_1$ and $S_2$ are respectively $D_1 = 30$ nm and $D_2 = 30$ nm. The width and height of the VG cavity are respectively $w_1 = 106$ nm and $h = 200$ nm, and the bottom width is $w_2 = 10$ nm. The diameter of the nanowire is $d = 12$ nm, and the distance from its bottom to the bottom of the VG cavity is $b = 5$ nm.

## II. WEAK EXCITATION LIMIT

First, we investigate the transmission response of this system in the weak excitation limit. Semiclassical theory is used to describe the transmission response of this system. The Bloch equation is the fundamental equation of semiclassical quantum optics. Based on the Bloch equation for the interaction of a single-mode electromagnetic field with a two-level system, the dipole transition equation for the interaction of the multimode electromagnetic field with the two-level system can be expressed as

$$\frac{d\rho_{eg}}{dt} = -(i\omega_A + \gamma)\rho_{eg} + iW \sum_m \frac{g_m}{\sqrt{\varepsilon_0 E_{vac}^m}} a_m(t) \quad (1)$$

where $\rho_{eg} = |e\rangle\langle g|$ is the dipole transition density matrix element, $\omega_A$ is the resonant frequency of the QEs in nanowire, $\gamma$ is the decay rate, $W = \rho_{ee} - \rho_{gg}$ is the population difference function, $E_{vac}^m$ is the vacuum field of the nth mode, and $g_m$ is the coupling rate. Changes in the light field are described by the classical method. According to the multimode interference coupled mode theory (MICMT) [27] and the Heisenberg operator equations [28], the following can be obtained

$$\frac{da_m}{dt} = -\left(i\omega_m + \frac{1}{\tau_{m0}} + \frac{1}{\tau_{m1}} + \frac{1}{\tau_{m2}}\right)a_m \quad (2)$$
$$- ig_m \sqrt{\varepsilon_0} E_{vac}^m \rho_{eg} + \kappa_{m1}s_{m,1+} + \kappa_{m2}s_{m,2+}$$

$$s_{1-} = -s_{1+} + \sum_m \kappa_{m1}^* a_m, \quad \kappa_{m1} = \sqrt{\frac{2}{\tau_{m1}}} \quad (3)$$

$$s_{2-} = -s_{2+} + \sum_m \kappa_{m2}^* a_m, \quad \kappa_{m2} = \sqrt{\frac{2}{\tau_{m2}}} e^{-i\phi_m} \quad (4)$$

$$s_{m,j+} = q_{mj}e^{i\varphi_{mj}} s_{j+}, \quad \sum_m q_{mj}e^{i\varphi_{mj}} = 1, \quad j = 1, 2 \quad (5)$$

Here, $a_m$ and $\omega_m$ are the normalized field amplitude and the resonant frequency of the mth resonant mode, respectively. $\tau_{m0}$ is the decay time of internal loss of the nth resonant mode in the resonator. $\tau_{m1}$ and $\tau_{m2}$ are the decay times of the coupling between the resonator and waveguides ($S_1$ and $S_2$), respectively. $\phi_m$ is the phase difference between the output port and the input port of the mth resonant mode. $s_{j\pm}$ are the field amplitudes in each waveguide (j = 1 and 2 , for outgoing (-) and incoming (+) from the resonator). Here, we primarily consider the interaction between the emitters and the kth mode whose resonant frequency is closest to the resonant frequency $\omega_A$ of the emitters, and ignore the interaction between the emitters and other modes whose resonant frequencies are far from the resonant frequency of the emitters (that is $g_m = 0, m \neq k$).

In the weak excitation limit, the QEs are predominantly in the ground state, this means $W \approx -1$. When $s_{2+} = 0$, the transmission coefficient of this hybrid system is

$$t = \frac{q_{k1}e^{i\varphi_{k1}}(i\Delta_A + \gamma)\kappa_{k1}\kappa_{k2}^*}{\left(i\Delta_k + \frac{1}{\tau_{k0}} + \frac{1}{\tau_{k1}} + \frac{1}{\tau_{k2}}\right)(i\Delta_A + \gamma) + g_k^2}$$
$$+ \sum_{m \neq k} \frac{q_{m1}e^{i\varphi_{m1}}\kappa_{m1}\kappa_{m2}^*}{i(\omega_m - \omega) + \frac{1}{\tau_{m0}} + \frac{1}{\tau_{m1}} + \frac{1}{\tau_{m2}}} \quad (6)$$
$$= t_k + \sum_{m \neq k} t_m$$

where, $\Delta_k = \omega_k - \omega$ and $\Delta_A = \omega_A - \omega$. If waveguides $S_1$ and $S_2$ of which the widths are equal are symmetric about the VG resonator, then $\tau_{m1} = \tau_{m2} = \tau_m$ and $\theta_{m1} = \theta_{m2}$, and the transmission coefficient formula is simplified as

$$t = \frac{2q_{k1}e^{i\varphi_k}(i\Delta_A+\gamma)}{\left(i\Delta_k\tau_k+2+\dfrac{\tau_k}{\tau_{k0}}\right)(i\Delta_A+\gamma)+\tau_k g_k^2}$$
$$+\sum_{m\neq k}\frac{2q_{m1}e^{i\varphi_m}}{i(\omega_m-\omega)\tau_m+2+\dfrac{\tau_m}{\tau_{m0}}}, \quad \varphi_m=\varphi_{m1}+\phi_m \tag{7}$$

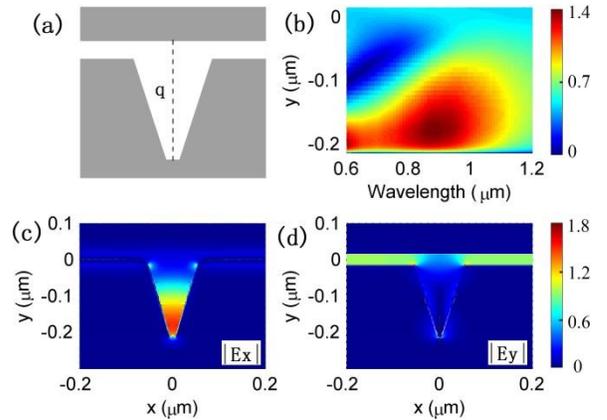

FIG. 2. (b) The distribution of electric field |E| on the dotted line q at different wavelengths. (c) and (d) The distribution of the electric field component |Ex| and |Ey| at the resonant wavelength of 851 nm.

Then, the corresponding transmittance of this hybrid plasmonic system is $T=|t|^2$.

We investigate the distribution of the fields in VG resonator without nanowire. The distribution of electric field $|E|$ on the dotted line q at different wavelengths is shown in Fig. 2(b). As can be seen from Fig. 2(b), the field enhancement in VG cavity can reach maximum at the wavelength of 851 nm, and the electric field distribution is mainly concentrated near the bottom of the VG cavity. The distribution of the electric field component $|Ex|$ and $|Ey|$ at the resonant wavelength of 851 nm are respectively given in Figs. 2(c) and 2(d), from which it can be seen that the electric field component in VG cavity is mainly $|Ex|$, and $|Ey|$ is almost zero. In order to obtain strong interaction between the electromagnetic fields and the emitters in nanowire, the nanowire should be placed near the bottom of the VG cavity, and the transition dipole moment of the QEs should be along the direction of x axis.

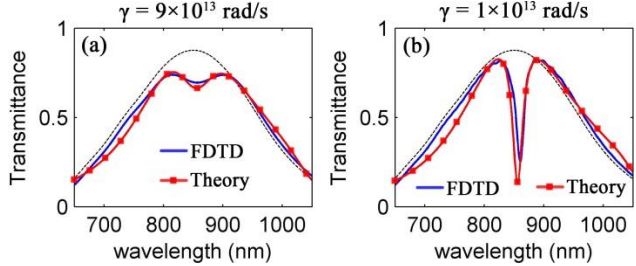

FIG. 3. The simulation results (blue lines) and theoretical results (red lines) for the transmittance of this hybrid system with the internal loss of the QEs are respectively (a) $\gamma=9\times10^{13}\ rad/s$ and (b) $\gamma=1\times10^{13}\ rad/s$. The black dotted line is the simulation result of the transmittance of this system without nanowire. Here, the Lorentz model parameters of the nanowire are taken as $\varepsilon_\infty=4.1$, $f=0.4$ and $\omega_L=2.17\times10^{15}\ rad/s$. The fitting parameters are respectively $q_{k1}=1$, $\varphi_k=0$, $\omega_k=2.17\times10^{15}\ rad/s$, $\omega_A=\omega_k$, $\tau_k=7\ fs$, $\tau_{k0}=48\ fs$, $q_\delta=0.3$, $\omega_\delta=1.78\times10^{15}\ rad/s$, (a) $\varphi_\delta=0.7\pi$, $\tau_\delta=42\ fs$, $\tau_{\delta0}=40\ fs$; (b) $\varphi_\delta=\pi$, $\tau_\delta=150\ fs$, $\tau_{\delta0}=100\ fs$.

For ease of analysis, the overall effect $\sum_{m\neq k} t_m$ of the mode whose resonant frequency is far form $\omega_A$ on the transmitance of this hybrid system is represented by a correction term $\delta=2q_\delta e^{i\varphi_\delta}/\left[i(\omega_\delta-\omega)\tau_\delta+2+\tau_\delta/\tau_{\delta0}\right]$. Thus, the transmission coefficient of this system is $t=t_k+\delta$. Simulation results (blue lines) and theoretical results (red lines) for the transmittance of this hybrid system with internal losses of the QEs are $\gamma=9\times10^{13}\ rad/s$ and $\gamma=1\times10^{13}\ rad/s$ are given in Figs. 3(a) and 3(b), respectively. Both of the simulation results and theoretical results show that the internal loss of the emitters in nanowire does not affect the coupling rate between the emitters and electromagnetic fields. By theoretical calculating, it can be obtained that the coupling rates in two cases both are $g_k=7\times10^{13}\ rad/s$.

## III. NON WEAK EXCITATION LIMIT

The previous discussions were carried out under weak excitation conditions. Rabi splitting occurs in the transmission spectrum of this system, indecating that the coupling between the emitters and VG resonator is strong coupling. When the excitation light intensity is sufficiently strong, the QEs in nanowire no longer remain in the weak excitation conditions, and the population difference function W(t) is a slowly varying function of time. So, the transmission coefficient of the system will also oscillate over time. The expression (Eqs. (6) and (7)) of the transmission coefficient are not applicable any more. Next, we will give the evolution of the transmission coefficient with time when this hybrid system is at resonance.

According to the Bloch equation and the Bloch vector (W,U,V)

$$W=\rho_{ee}-\rho_{gg},\quad U=\frac{1}{2}(\rho_{eg}+\rho_{ge}),\quad V=\frac{1}{2i}(\rho_{eg}-\rho_{ge}) \tag{8}$$

If the emitters (for which the internal loss is neglected) are in the ground state at the initial time, it can be obtained that

$$\frac{dW}{dt}=2\Omega_R(t)\cdot\left[U\sin(\Delta_A t)-V\cos(\Delta_A t)\right] \tag{9}$$

$$\frac{dU}{dt}=-\frac{\Omega_R(t)}{2}\cdot W\sin(\Delta_A t) \tag{10}$$

$$\frac{dV}{dt}=\frac{\Omega_R(t)}{2}\cdot W\cos(\Delta_A t) \tag{11}$$

where, $\Omega_R=-\boldsymbol{\mu}_1\cdot\boldsymbol{\psi}(\mathbf{r})|a|/(\hbar\sqrt{\varepsilon_0})=\chi|a|$. When the system is at resonace ($\Delta_A=0$), the above equation is simplified as

$$\frac{dW}{dt}=-2\Omega_R(t)\cdot V \tag{12}$$

$$\frac{dV}{dt}=\frac{\Omega_R(t)}{2}\cdot W \tag{13}$$

The initial condition is $W(0) = -1$, $\dot{W}(0) = 0$, $V(0) = 0$. Then, the solutions of Eqs. (12) and (13) are

$$W(t) = -\cos\left(\int_0^t \Omega_R(t)dt\right) \quad (14)$$

$$V(t) = -\frac{1}{2}\cdot\sin\left(\int_0^t \Omega_R(t)dt\right) \quad (15)$$

If the total number of QEs in nanowire is N and the mode volume of the resonantor is $V_{eff}$, the number density of QEs in nanowire is $n = N/V_{eff}$. Ignoring the interaction between the QEs, the output power density of the nanowire is

$$\begin{aligned}p_{nw} &= -n\hbar\omega_A \cdot \frac{1}{2}\frac{dW}{dt} \\ &= -n\hbar\omega_A \cdot (\Omega_R/2)\sin\left(\int_0^t \Omega_R(t)dt\right)\end{aligned} \quad (16)$$

Eq. (16) shows that the coupling between the electromagnetic field and the emitter causes the output power density of the nanowire to behave as an oscillating form. The oscillating output power in turn causes the field amplitude in the resonator to oscillate over time. A feedback system is formed between the plasmonic resonator and the QEs in nanowire. We assume that $a_{nw}$ is the extra field amplitude induced by the output power of the nanowire, $a_0$ is the field amplitude in plasmonic resonator without QEs in nanowire, $a$ is the total field amplitude. According to the coupled mode theory (CMT), the field in plasmonic resonator satisfies the following relationship

$$\frac{da_0}{dt} = -\left(i\omega_k + \frac{1}{\tau_{k0}} + \frac{1}{\tau_{k1}} + \frac{1}{\tau_{k2}}\right)a_0 + \kappa_{k1}s_{1+} + \kappa_{k2}s_{2+} \quad (17)$$

$$s_{1-} = -s_{1+} + \kappa_{k1}^*(a_0 + a_{nw}) = -s_{1+} + \kappa_{k1}^* a \quad (18)$$

$$s_{2-} = -s_{2+} + \kappa_{k2}^*(a_0 + a_{nw}) = -s_{2+} + \kappa_{k2}^* a \quad (19)$$

For symmetric systems, $\tau_{k1} = \tau_{k2} = \tau_k$, and if $s_{2+} = 0$, then the solution of Eq. (17) is

$$a_0 = \frac{\kappa_{k1}s_{1+}}{i(\omega_k - \omega) + \frac{2}{\tau_k} + \frac{1}{\tau_{k0}}} \quad (20)$$

and

$$s_{2-} = \kappa_{k2}^* a \quad (21)$$

When the system is at resonance ($\omega_k - \omega = 0$), there is

$$\kappa_{k1}s_{1+} = \left(\frac{2}{\tau_k} + \frac{1}{\tau_{k0}}\right)a_0 \quad (22)$$

According to power conservation

$$|s_{1-}|^2 + |s_{2-}|^2 = |s_{1+}|^2 + p_{nw} - \frac{2}{\tau_{k0}}|a|^2 \quad (23)$$

and Eqs. (18) and (21), it can be obtained that

$$\begin{aligned}2\times\left(\frac{2}{\tau_k} + \frac{1}{\tau_{k0}}\right)|a|^2 &- s_{1+}\kappa_{k1}a^* - s_{1+}^*\kappa_{k1}^*a \\ &= -\frac{\chi}{2}\sin\left(\int_0^t \Omega_R(t)dt\right)\cdot n\hbar\omega_A \cdot |a|\end{aligned} \quad (24)$$

where $\chi = \mathbf{\mu}_1 \cdot \mathbf{\psi}(\mathbf{r})/(\hbar\sqrt{\varepsilon_0})$. In order to facilitate the calculation, we take $\kappa_{k1}s_{1+}$ as a real number. In this case, $a_0$ and $a$ both are real numbers when the system is at resonance, then Eq. (24) is simplified as

$$\left(1 + \frac{\tau_k}{2\tau_{k0}}\right)(a - a_0) + \frac{\chi\tau_k}{8}\sin\left(\int_0^t \Omega_R \cdot dt\right)\cdot n\hbar\omega_A = 0 \quad (25)$$

The transmittance of the system is

$$T = \left|\frac{s_{2-}}{s_{1+}}\right|^2 = \left|\frac{2}{i(\omega_k - \omega)\tau_k + 2 + \frac{\tau_k}{\tau_{k0}}}\right|^2 \cdot \left|\frac{a}{a_0}\right|^2 \quad (26)$$

$$= T_0 \left|\frac{a}{a_0}\right|^2$$

$T_0$ is the transmittance of the system when there is no QEs in nanowire. The oscillating field amplitude (a) can be obtained from Eq. (25). And Eq. (26) shows that the oscillating field amplitude causes the transmitance of the system to oscillate over time. Based on Eqs. (25) and (26), we will investigate the influences of the incident light intensity and QE number density on the transmission response for this hybrid system.

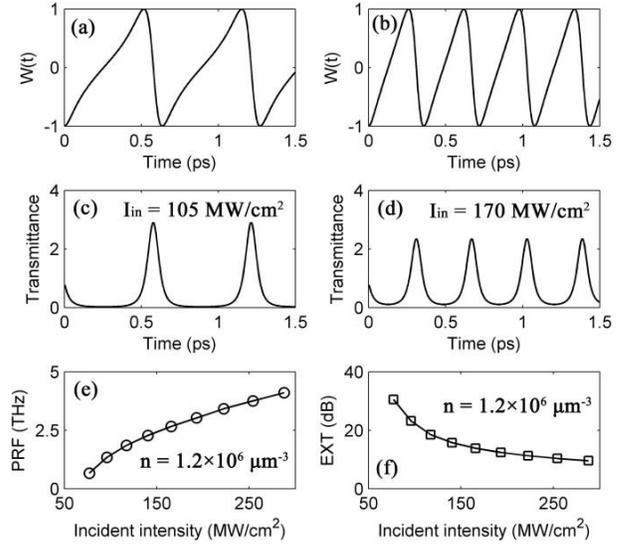

FIG. 4. (a)-(d) The curves for the population difference function and transmittance changing with time at different incident light intensity $I_{in} = 105\ MW/cm^2$ and $I_{in} = 170\ MW/cm^2$ when the number of emitters in unit volume remains a constant $n = 1.2\times 10^6\ \mu m^{-3}$. (e) and (f) The curves for the repetition frequency and extinction ratio of the pulses changing with incident light intensity when the number density of QEs is $n = 1.2\times 10^6\ \mu m^{-3}$. The incident light intensity is expressed as $I_{in} = c|a_0|^2/(n_{eff}T_0)$, c is the speed of light in free space, $n_{eff} = 1.62$ is the effective refractive index of the waveguide, $T_0 = 0.87$. The nanowire is placed at the position of $\psi(\mathbf{r}) = 1$.

When the QE number density remains a constant $n = 1.2 \times 10^6 \ \mu m^{-3}$, the results for the population difference function and transmittance over time at different incident light intensities are shown in Figs. 4(a)-4(d), revealing that the incident light intensity has a significant effect on both the pulse repetition frequency (PRF) and the extinction ratio. Curves for the repetition frequency and extinction ratio as a function of incident light intensity are given in Figs. 4(e) and 4(f). The repetition frequency of the pulses increases from 0.67 THz to 4.1 THz as the incident light intensity increases from 77 MW/cm$^2$ to 288 MW/cm$^2$, but the extinction ratio $\left(EXT = 10 \cdot \lg\left(T_{max}/T_{min}\right)\right)$ decreases from 31 dB to 9.6 dB. This is because the greater the incident light intensity, the smaller the influence of output power of the nanowire on the transmission power. In Fig. 4(c), the pulse width is 83 fs, and the repetition frequency is 1.67 THz. This result indicates that it is feasible to generate femtosecond pulses with THz repetition frequencies in plasmonic resonator systems using a continous wave input.

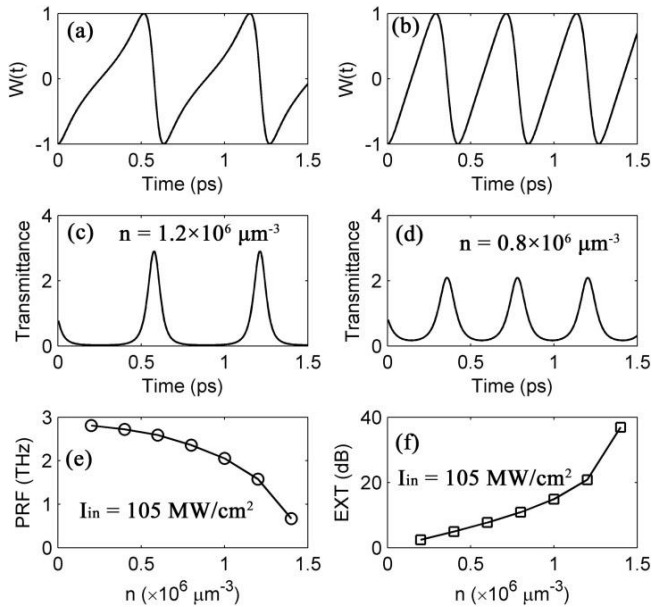

FIG. 5. (a)-(d) The curves for the population difference function and transmittance changing with time at $n = 1.2 \times 10^6 \ \mu m^{-3}$ and $n = 0.8 \times 10^6 \ \mu m^{-3}$ when the incident light intensity is $I_{in} = 105 \ MW/cm^2$. (e) and (f) The curves for the repetition frequency and extinction ratio of the pulses changing with the number density of QEs when the incident light intensity is $I_{in} = 105 \ MW/cm^2$. The nanowire is placed at the position of $\psi(\mathbf{r}) = 1$.

When the incident light intensity is a constant $I_{in} = 105 \ MW/cm^2$, results for the population difference function and the transmittance over time at $n = 1.2 \times 10^6 \ \mu m^{-3}$ and $n = 0.8 \times 10^6 \ \mu m^{-3}$ are given in Figs. 5(a)-5(d), respectively. By contrast, it can be found that the larger the number density of QEs, the closer the population difference function curve is to a zigzag of which the front slope is gentle and the back slope is steep, and the closer the transmittance curve is to pulse form. Moreover, the repetition frequency and extinction ratio of the pulses can alse be controlled by the QE number density. Figures 5(e) and 5(f) show the results for changes in the repetition frequency and extinction ratio of the pulses as a function of the QE number density. As n increases from $n = 0.2 \times 10^6 \ \mu m^{-3}$ to $n = 1.4 \times 10^6 \ \mu m^{-3}$, the repetition frequency of the pulses decreases from 2.8 THz to 0.67 THz, whereas the extinction ratio increases from 2.4 dB to 37 dB.

## IV. CONCLUSION

In summary, we have investigated the transmission response of a hybrid system composed of a VG plasmonic resonator coupled to a nanowire embedded with two-level QEs. Femtosecond pulses with THz repetition frequency can be obtained by the strong coupling between the QEs and the plasmonic resonator in the case of continuous wave input. The repetition frequency and extinction ratio of the pulses can be freely controlled by the incident light intensity and QE number density. The pulse width can be modulated below 100 fs, and the extinction ratio can also reach very high values. This provides a method for generating narrow pulsed lasers with ultrahigh repetition frequency in plasmonic resonator system, which has potential applications for generating optical clock signals at the nanoscale.

## ACKNOWLEDGMENTS

This work was supported by the National Key R&D Program of China (2016YFA0301300); National Natural Science Foundation of China (NSFC) (11574035, 11374041, 61571060, 11404030).